# Tuning nanoscale adhesive contact behavior to a near ideal Hertzian state via graphene coverage

Yongchao Chen[1], Zhizi Guan[1], Wei Yang[1], Yongtao Yao[2], Hailong Wang[1, *]

[1] CAS Key Laboratory of Mechanical Behavior and Design of Materials, Department of Modern Mechanics, University of Science and Technology of China, Hefei 230027, China

[2] National Key Laboratory of Science and Technology on Advanced Composites in Special Environments, Harbin Institute of Technology, Harbin, Heilongjiang 150080, China

*E-mail: hailwang@ustc.edu.cn

**ABSTRACT**

We carry out molecular statics (MS) simulations to study the indentation process of Pt (111) surfaces using an indenter with the radius of 5-20 nm. The substrate and indenter surfaces are either bare or graphene-covered. Our simulations show that the influence of the adhesion between the bare substrate and indenter tip can be significantly reduced by decreasing the adhesion strength and adhesion range between the atoms on the substrate and indenter, or by enhancing the substrate stiffness. Our results suggest that the elastic response of the substrate exhibits weaker adhesion after the coating of graphene layers on either side of the contacting interface, which is attributed to the weak interaction between the graphene layers. Based on these principles obtained for the bare substrate, the nanoscale contact behavior of the substrate can be tuned into a near-ideal Hertzian state by increasing the number of graphene layers, applying pre-strains to graphene on substrate, or using large indenters. Therefore, even for strong adhesion between tip, graphene, and substrate, mechanical properties of substrate can be determined by nanoindentation using graphene coverage. Our research provides theoretical guidance for designing adhesion-less coatings for AFM probes and MEMS/NEMS systems.

## 1. Introduction

Graphene (Gr) is a two-dimensional (2D) material made of a single layer of carbon atoms [1] that possesses several exotic properties that make it uniquely suited for use as a protective coating and superior lubricant for electrical contacts [2, 3]:an ultimate tensile strength of 130 GPa [4], an ultralow surface energy of 46.7 mJ/m$^2$ [5, 6], and an atomically-thin layer [7]. As a result, remarkable achievements such as layer-dependent transient friction strengthening [8], friction tuning via in-pane straining [9], and nanowear reduction by suppressing the contact pressure fluctuation [10] have been obtained in the exploration of the important low-friction and anti-wear properties of graphene. In addition to the graphene effects in lubrication, molecular dynamics (MD) simulations of nanoindentation show that a graphene coating can make metallic substrates more resistant to plastic deformation and considerably improve the load-bearing capacity; this has been attributed to atomic-scale surface roughness smoothing [11, 12], homogenization of the contact stress distribution beneath the tip [13] and interactions between the dislocations and graphene coating interface [14].

Most of the studies of friction and nanoindentation have focused on the lateral friction behaviors [8, 9, 15, 16] and plastic deformation [11-13] but have not performed detailed investigations of the behaviors during the initial elastic contact stage. However, in fact, elastic contact is not only an inevitable process prior to the friction and plastic stage, but is also a completely different state for the graphene protective mechanism. Recently, it was reported that the experimental elastic load-displacement curves from a sub-nanometer indentation on the graphene-coated Cu film are in a closer agreement with the Hertz model compared to bare Cu [17, 18]. Meanwhile, MD simulations have also revealed that graphene regulates the interaction between the indenter tip and the substrate and yields a Hertz-type contact (with the modulus of the bulk substrate) due to the elimination of interatomic adhesion [11, 18]. Simultaneously, tremendous research efforts have been devoted to the studies of the adhesive contact behavior between various materials such as the theoretical description of adhesive contact [19, 20], material transfer due to adhesion [21] and different factors influencing the adhesion strength [22, 23]. Hence, a well-established theory of this modification-by-graphene phenomenon will open a new route for the research on adhesive contact. However, many factors that influence this regulating effect such as the number of graphene layers, in-plane pre-strain in graphene, and the size of indenter, are generally neglected, even though all of these may exert strong effects on the elastic response to nanoscale adhesive contact. Due to the importance of surface adhesion and the contact behavior of graphene-coated surfaces, a better insight into graphene's adhesion and contact properties is highly necessary.

In this work, through molecular statics (MS) simulations, we show that the elastic contact behavior of a metallic substrate exhibits a high load bearing capacity as well as a near ideal Hertzian contact that can be achieved by coating graphene on either side of the contact interface (upper spherical or lower flat surfaces). We analyze the atomic-scale microstructure in the contact region between the tip and bare substrate, and find that the adhesion can be mitigated by reducing the adhesion strength and adhesion range of the interaction between influence atoms on influence tip and influence substrate, or enhancing the stiffness of influence substrate surface. More specifically, the adhesion between the tip and graphene-coated substrate can be significantly reduced by increasing the covered graphene layers or applying pre-strains to the graphene monolayer, further regulating the elastic response and yielding a Hertz-type contact. Moreover, the regulation of the elastic contact behavior of a graphene-coated substrate can be enhanced by increasing the indenter size due to the reduction of the relative adhesive area on the indenter, particularly for high adhesion of the tip to the substrate. As demonstrated in previous studies [11, 12, 17, 18, 24, 25], this adhesion-less contact behavior regulated by graphene layers enables the measurement of the elastic modulus for various substrate materials via fitting the load-displacement curves extracted from nanoindentation to the Hertz theory. Aside from fundamental interest, proper tribological engineering that takes adhesion into account is crucial for the development of reliable MEMS devices such as the digital mirror and MEMS switches [26, 27] for which the transfer between the contact and materials is generally an important issue. Hence, our investigation provides theoretical guidance for designing adhesion-less coatings for the AFM probes and MEMS/NEMS systems [28].

## 2. Methods

MS simulations in LAMMPS [29] are carried out to model the nanoindentation process on a Pt substrate with a rigid semispherical diamond tip. The tips are moved in steps of 0.05 Å, after which the system is relaxed to the next local minimum until atomic forces drop below $10^{-4}$ eV/Å. To eliminate the influence of atomic-scale geometry, the tips are produced by displacing the atoms of a (111) face-centered cubic (fcc) slab normal to the slab surface such that surface atoms precisely follow the surface of a sphere [11]. The radii of the bare or graphene-layers-wrapped tips are varied from 5 to 20 nm to study the size effects. The substrates are also bare or coated with graphene layers, periodically repeated cubic Pt crystals in their face centered cubic (fcc) ground-state structure with the (111) surfaces exposed toward the indenter. Graphene layers are placed on the surfaces with the graphene <10> direction parallel to the Pt <110> direction. One layer of atoms at the bottom of the surface is fixed during the calculations to anchor the solid within the simulation cell. To minimize the influence of periodic indenter images, the side length and height of the substrate are chosen to be roughly five and two times the indenter radius, respectively. Additional simulations for substrates with larger dimensions are carried out and suggest that our results do not depend on the size of substrate.

The embedded atom method (EAM) potential [30] is used to model the Pt-Pt interaction within the substrate. The C-C interaction within the same graphene sheet is modeled using the second-generation empirical reactive bond-order (REBO2) potential [31] with a modified cutoff scheme [32]. To study the effect of substrate stiffness, the EAM [33-34] potentials are applied for the Pb and Al substrates, respectively. The truncated

Lennard-Jones (LJ) potential is used for the atomic interactions between the tip, graphene, and substrate,

$$U(r) = \begin{cases} 4\varepsilon\left(\left(\frac{\sigma}{r}\right)^{12} - \left(\frac{\sigma}{r}\right)^6\right) - 4\varepsilon\left(\left(\frac{\sigma}{r_c}\right)^{12} - \left(\frac{\sigma}{r_c}\right)^6\right), r < r_c \\ 0, r \geq r_c \end{cases} \quad (1)$$

where $\varepsilon$ and $\sigma$ are parameters for LJ potential, $r_c = 2.5\sigma$ is the cutoff radius. We set $\varepsilon_{Tip-Sub} = 0.0112\ eV$ and $\sigma_{Tip-Sub} = 0.297$ nm for the interaction between the tip and the Pt substrate, $\varepsilon_{Gr-Tip} = 0.00284$ eV and $\sigma_{Gr-Tip} = 0.34$ nm for the interaction between the tip and graphene, $\varepsilon_{Gr-Sub} = 0.0112$ eV and $\sigma_{Gr-Sub} = 0.297$ nm for the interaction between the Pt substrate and graphene [35]. $\varepsilon_{Gr} = 0.00284$ eV and $\sigma_{Gr} = 0.34$ nm are the parameters for the interaction between different graphene layers. To investigate the effects of adhesion strength and adhesion range on the nanoscale contact behavior, we tune the depth of the potential well ($\varepsilon$) and the equilibrium distance ($\sigma_0 = 2^{1/6}\sigma$). For bare substrate, we set $\varepsilon_{Tip-Sub} = 0.0224$ eV and $\varepsilon_{Tip-Sub} = 0.0056$ eV for strong and weak adhesion, respectively, and $\sigma_{Tip-Sub} = 0.1485$ and $\sigma_{Tip-Sub} = 0.594$nm for short-range and long-range adhesion, respectively. For the graphene-coated substrate, to optimally model extreme conditions, we artificially enlarge the depth of the potential well for strong tip-graphene-substrate adhesion. Therefore, four typical adhesion cases are considered for nanoindentation on the graphene-covered substrate in this work, namely, normal adhesion ($\varepsilon_{Gr-Tip} = 0.0112$ eV, $\varepsilon_{Gr-Sub} = 0.0112$ eV), strong Gr-Sub adhesion ($\varepsilon_{Gr-Tip} = 0.0112$ eV, $\varepsilon_{Gr-Sub} = 0.112$ eV), strong Gr-Tip adhesion ($\varepsilon_{Gr-Tip} = 0.112$ eV, $\varepsilon_{Gr-Sub} = 0.0112$ eV), and strong adhesion ($\varepsilon_{Gr-Tip} = 0.112$ eV, $\varepsilon_{Gr-Sub} = 0.112$ eV).

## 3. Results and Discussion

### 3.1. Influential factors for nanoscale adhesive contact on bare substrate

We start with indenting the bare substrate and analyzing the force versus displacement curves from these simulations to study the effects of tip-substrate adhesion and substrate stiffness on the contact behavior. In this regard, Hertz theory predicts that the force $P$ due to a displacement $h$ on a sphere of radius R is given by,

$$P = (4E^*R^2/3)\ (h/R)^{3/2} \quad (2)$$

where the contact modulus $E^* = E/(1 - \nu^2)$ for a rigid and isotropic substrate. To quantify the deviation from the Hertz prediction, the exponent $N$ and contact modulus $E^*$ of a specific contact behavior is determined by fitting to $P/(4E_R^*R^2/3) = (E^*/E_R^*)\ (h/R)^N$ extracted from the load-displacement curve, where $E_R^* = 207$GPa is the fitted elastic modulus when employing a purely repulsive interaction. Fig. 1 summarizes the plots of force $P$ on the indenter as a function of the scaled penetration depth $h/R$ obtained for the indenter radius $R = 5$ nm on bare substrate. Fig. 1a shows that in the elastic regime, the purely repulsive contact behavior is highly consistent with Hertz theory with a fitting exponent $N = 1.49$ and $E^* = 205$GPa (red curve), consistent with $E^* = 202 \text{GPa}$ for the EAM Pt potential [30]. In Fig. 1a, the negligible deviation from the Hertz prediction ($N = 1.50$) is due to the morphological difference between a stepped tip and a strictly spherical tip. For comparison, the force-displacement curve is also plotted for the normal tip-substrate adhesion in all panels ($N = 1.33$ and $E^* = 190 \text{GPa}$ for black curves), in which the deviation from Hertz theory is apparent due to the attractive contribution of the Lennard-Jones potential. This result indicates that with a purely repulsive interaction, nanoindentation can follow Hertz theoretical prediction even in the absence of graphene.

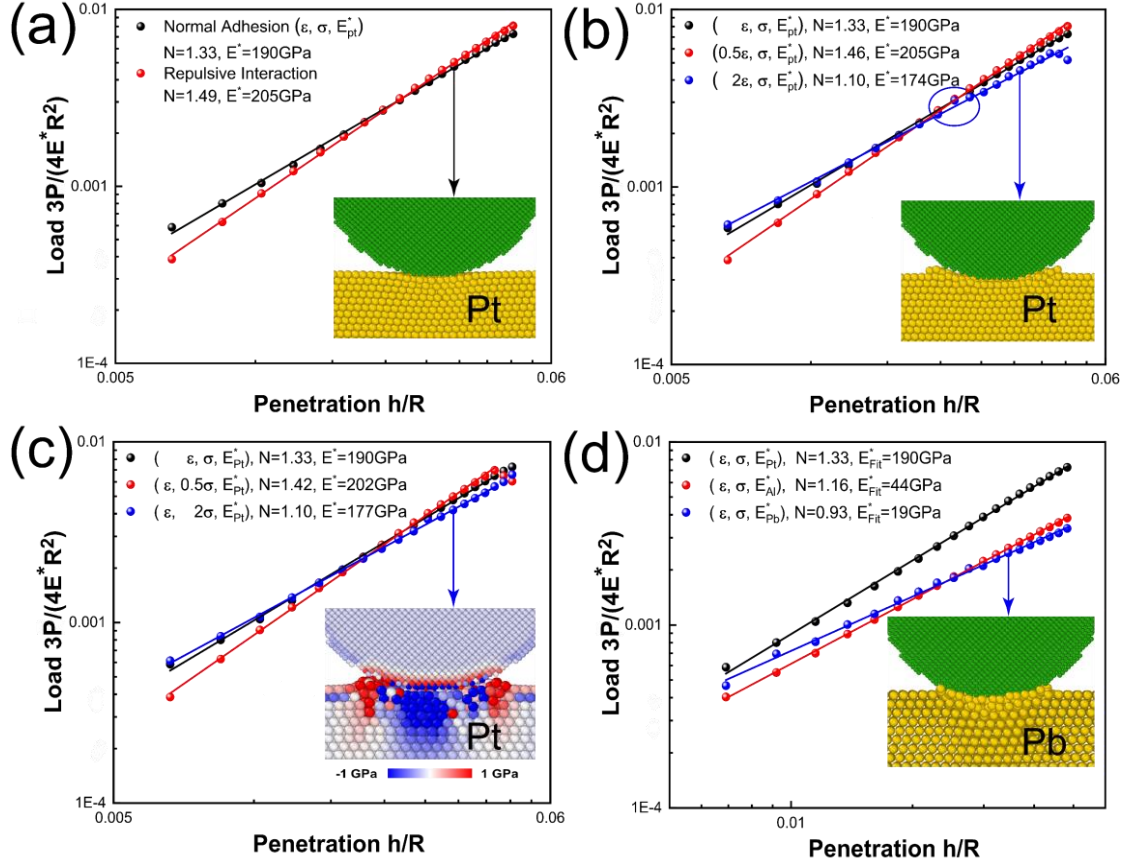

**Figure 1.** Factors that influence nanoscale adhesive contact on the bare substrate. (a) A near-ideal Hertz contact behavior when employing a purely repulsive interaction, compared with a normal interaction case ($\varepsilon_{Tip-Sub} = 0.0112$ eV, $\sigma_{Tip-Sub} = 0.297$ nm, $E^* = E^*_{Pt}$). (b-d) The effects of (b) adhesion strength ($\varepsilon_{Tip-Sub} = 0.0224, 0.0112, 0.0056$ eV), (c) adhesion range ($\sigma_{Tip-Sub} = 0.1485, 0.297, 0.594$ nm), and (d) substrate stiffness ($E^* = E^*_{Pt}, E^*_{Al}, E^*_{Pb}$) on the degree of deviations from the Hertz prediction. Plots of force $P$ vs. scaled penetration depth $h/R$ for the indenter radii of $R = 5$ nm are depicted in each subfigure. (insets) The snapshots (green for the tip atoms and yellow for the substrate atoms) indicate stronger deviations for (b) strong adhesion strength, (c) long adhesion range, and (d) soft substrate contributed to more substrate atoms transferring to the tips. The inset in (c) illustrates the stress distribution during contact where the color of each atom represents the atomic stress $\sigma_{zz}$ component. For consistency, the color scheme for the atomic plots is preserved in the remaining figures. The indenter radius of $R = 5$ nm is maintained in all of atomic simulations unless mentioned otherwise.

To further investigate the effect of adhesion between the tip and the substrate, we systematically tune the adhesion strength, range of atomic interaction and substrate stiffness to compare with the normal adhesion. Fig. 1b shows that compared with the normal adhesion ($N = 1.33$ and $E^* = 190$GPa for black curve), weak adhesion leads to a smaller deviation ($N = 1.46$ and $E^* = 205$GPa for red curve) from the Hertz prediction. Furthermore, force fluctuation (blue circle in Fig. 1b) can arise due to the instantaneous adhesion of Pt atoms to the tip when the adhesive force is strong enough. The inset of Fig. 1b clearly shows that atoms transfer to the tip during the contact process, which is not observed in the snapshot for the normal adhesion (inset, Fig. 1a). The influence of the equilibrium distance is illustrated in Fig. 1c that shows that a short-range adhesive interaction gives rise to less deviations ($N = 1.42$ and $E^* = 202$GPa for red curve) for the contact forces.

The inset of Fig. 1c shows the specific distribution of atomic stress σ$_{zz}$, in which only the atoms of the outer sphere and near the contact region are affected by the apparent adhesive forces (red colored). It is clear that this phenomenon is caused by the restrictions from the force range between the indenter and the substrate. We also note that the large force drop in some curves in Fig. 1 indicates the plastic deformation in the substrate. Additional MS simulations are performed to compare the adhesive contact behaviors of the substrates with various stiffness values E$^*$. Here, we change the substrate materials into Pb ($E^*_{Pb} = 39$GPa, fcc lattice structure) and Al ($E^*_{Al} = 78$ GPa, fcc lattice structure), respectively. Fig. 1d illustrates that, compared to softer Pb and Al substrates, a stiffer Pt substrate suffers less deviation from adhesion and exhibits a contact behavior closer (black curve) to that predicted by the Hertz theory. This can be explained as due to the atoms at the contact area being strongly constrained by the surrounding atoms and cannot adhere to the tip. As a result, the hypothesis that adhesion can impose a deviation from ideal Hertz contact behavior is corroborated and more specifically, this deviation has three main characteristics, i.e., lower $N$ values, lower fitted contact modulus $E^*$, and greater force fluctuation.

Therefore, there are three main factors affecting the adhesion, namely adhesion strength, adhesion range, and substrate stiffness that provide effective guidance for adhesion mitigation. Naturally, this gives rise to the question on how to minimize or even eliminate adhesion effects during nanoindentation.

### 3.2. Reduce adhesion strength by coating graphene layers

Recent studies [15, 16, 18] have revealed that coating the substrate with graphene can mitigate the adhesion between the substrate and the tip. To elucidate the underlying mechanisms of the effects of the graphene coating on nanoindentation, MS simulations are conducted on a Pt substrate with graphene layers. Moreover, a systematic analysis is performed to investigate the regulation of the elastic response of graphene-layers-covered substrate to nanoindentation based on the principles obtained for bare substrate.

*Graphene monolayer coating:* We coat a monolayer of graphene onto the substrate and simulate the indentation process of a diamond tip contacting a Pt substrate (inset, Fig.

2a). As the distance between the substrate and the tip increases due to the isolation of graphene, the adhesion on the tip is significantly reduced and the transfer of the substrate atoms to the tip is also substantially constrained; thus, this appears to be an ideal method to eliminate adhesion. Specifically, the contact response of the graphene-coated substrate shows close-to-optimal contact behavior ($N = 1.41$ and $E^* = 198\text{GPa}$ for black curve in Fig. 2a), in which both the graphene-substrate and graphene-tip interactions are weak Pt-C adhesion. Generally, graphene shows weak adhesion to other materials, thus regulating the contact behavior to follow the Hertz prediction. However, in some cases, graphene may unavoidably have strong adhesion with the tip or the substrate such as the graphene-mica interaction [15, 36, 37] and interaction with oxidized graphene [38]. Therefore, it is urgently needed to study nanoscale contact in the presence of strong adhesion for either tip-graphene or graphene-substrate interaction.

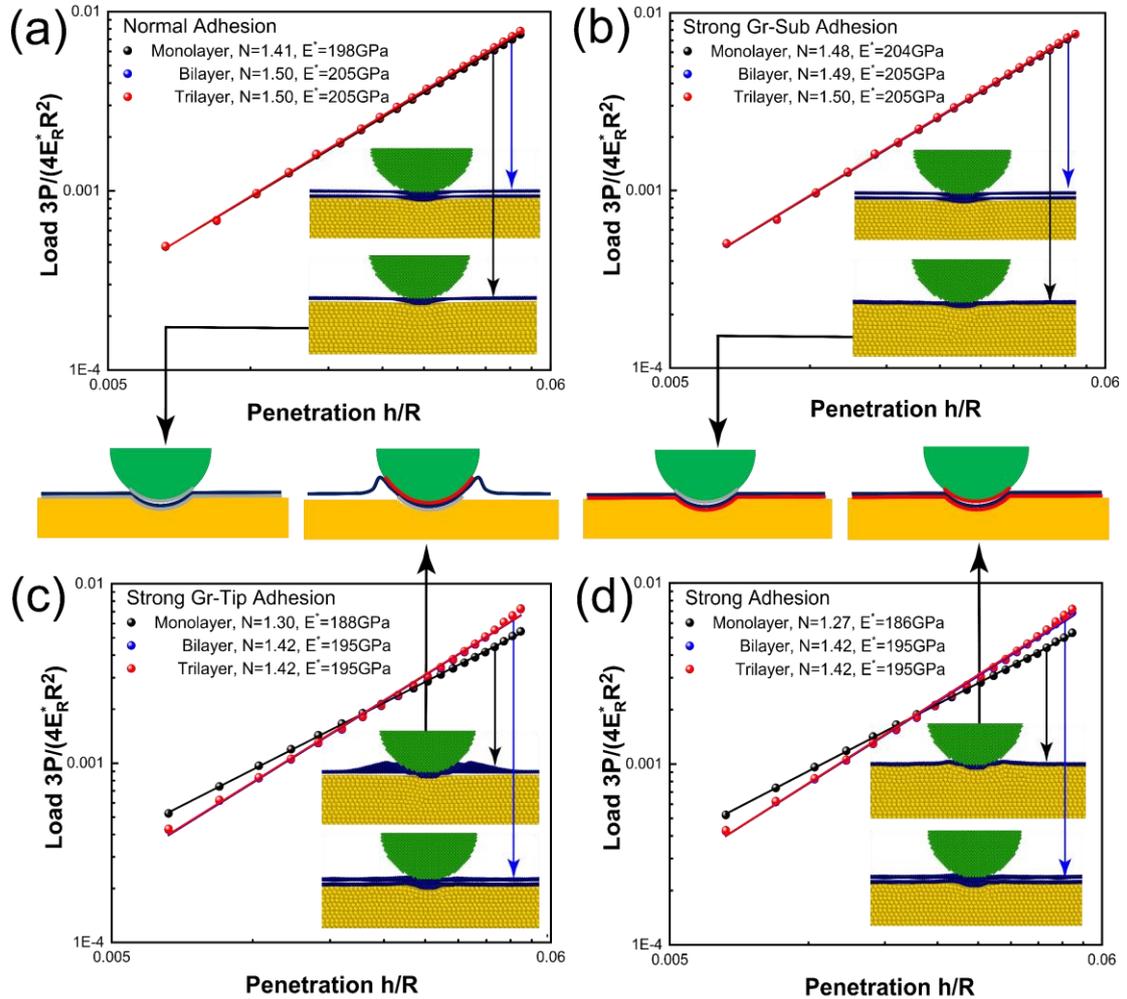

**Figure 2.** Influence of the number of layers of the graphene coatings on the degree of deviation from the Hertz prediction for typical interactions between the tips, graphene layers, and substrates: (a) normal adhesion ($\varepsilon_{Gr-Tip} = 0.0112$ eV, $\varepsilon_{Gr-Sub} = 0.0112$ eV), (b) strong Gr-Sub adhesion ($\varepsilon_{Gr-Tip} = 0.0112$ eV, $\varepsilon_{Gr-Sub} = 0.112$ eV), (c) strong Gr-Tip adhesion ($\varepsilon_{Gr-Tip} = 0.112$ eV,

$\varepsilon_{Gr-Sub} = 0.0112$ eV), and (d) strong adhesion ($\varepsilon_{Gr-Tip} = 0.112$ eV, $\varepsilon_{Gr-Sub} = 0.112$ eV). Plots of force $P$ vs. scaled penetration depth $h/R$ for the graphene monolayer, bilayer, and trilayer coatings on the substrate are depicted in each subfigure. The insets show the snapshots of nanoindentation on the monolayer and bilayer graphene-coated substrates. Schematic illustrations of the tip-graphene-substrate adhesion are also presented for the graphene monolayer case, where the gray and red regions indicate normal ($\varepsilon_{C-Pt}$) and strong adhesion ($10\varepsilon_{C-Pt}$), respectively.

Figs. 2b-c illustrate two corresponding cases when strong adhesion is applied to graphene-substrate adhesion and graphene-tip adhesion, respectively. It is demonstrated that the contact behavior is highly consistent with the Hertz contact prediction in the strong Gr-Sub adhesion case ($N = 1.48$ and $E^* = 204$GPa for black curve in Fig. 2b), whereas the graphene apparently loses its regulation effect in the strong Gr-Tip adhesion case ($N = 1.30$ and $E^* = 188$GPa for black curve in Fig. 2c). Fig. 2b (inset) schematically illustrates that a strong graphene-substrate adhesion will not lead to force deviations because the tip force is not directly related to the graphene-substrate interaction. Strong graphene-substrate adhesion can even contribute to the regulation of the contact behavior by constraining graphene deformation and the interfacial contact between graphene and the tip, thus weakening graphene-tip adhesion [15, 16]. Consistently, it was demonstrated by Niu et al. [39] and Zhou et al. [40] that normal graphene-tip adhesion can be neglected when fitting the results of FEM and MD simulations to the results of the indentation experiment if a strong graphene-substrate adhesion exits. As shown in Fig. 2c (inset), a proportion of graphene tightly adheres to the tip and the rest of the graphene covering the substrate significantly impacts the movement of the tip, thus leading to large deviations in the loading force. For the strong Gr-Tip adhesion case, the high out-of-plane flexibility of graphene coupled with strong tip–graphene adhesion causes local puckering near the contact edge.

The force-displacement curve for the strong adhesion case presents a strong deviation from the Hertz contact behavior ($N = 1.27$ and $E^* = 186$GPa for black curve in Fig. 2d) that is due to the observed firm attachment of both sides of interface by graphene that acts as the intermediate for the adhesion force (inset, Fig. 2d). It should also be attributed to the fact that force fluctuation does not appear in the strong Gr-Tip adhesion case even if strong adhesion exists between the tip and the contact surface. This is attributed to high in-plane stiffness of graphene providing an additional restriction to the single atoms' transfer behavior. We conclude that the effectiveness of graphene in modifying the contact response also strongly relies on its unique properties, i.e., high in-plane stiffness [4] and low surface energy [5, 6]. This is also the reason why other coatings such as oxide layers lack this ability.

*Multilayer graphene coatings:* To further regulate adhesive contact behaviors for the monolayer system in the strong adhesion case, we increase the thickness of the graphene coating to bilayer and trilayer (blue and red curves in Fig. 2). It has been reported that the friction of graphite and other lamellar materials increases with decreasing number of atomic layers [15, 16, 38]. Fig. 2d shows that the fitted exponent (1.42 and $E^* = 195$GPa for blue curve) for the load-displacement curve of bilayer coating of graphene is much closer to 1.50 than that of the monolayer coating (1.27 and $E^* = 186\text{GPa}$ for black curve), but the small difference is due to the lateral constraint from the surface graphene layer. The schematic indicates that the transmission of adhesion between the tip and the substrate is blocked due to the weak interlayer interaction of graphene layers. As we further increase the thickness of the graphene coating to trilayer, the fitted exponent for the load-displacement curve is also N = 1.42 and $E^* = 195\text{GPa}$, same as that of the bilayer coating. For a strong tip-graphene and graphene-substrate interaction, the layer-dependent regulation can lead to the conclusion that the dominant effects in the change from monolayer to bilayer are due to the adhesion blocking by weak graphene interlayers interaction, but the same performance of the bilayer and multilayer is the result of the lateral constraint from graphene on the top surface. We note that this trend can be further supported by the experimental results for graphene-coated nanoindentation obtained by Suk et al. [41] who found that multilayer coating has lower adhesion energy and force gap. We then apply the multilayer coverage system for the strong Gr-Tip adhesion case (Fig. 2c) and observe a contact response nearly identical to that observed in the strong adhesion case. This corroborates the abovementioned finding that the graphene/substrate interaction cannot provide a substantial contribution to the tip force when a multilayer is applied.

*Graphene coating on tip*: Considering that in the above-mentioned strong Gr-Tip adhesion case where some of the graphene atoms tightly adhere to the tip and lead to a severe deviation of the contact behavior, we naturally propose a method to coat a graphene monolayer directly onto the tip rather than the substrate [36, 42, 43]. Fig. 3 (inset) schematically elucidates the MS setup for the graphene coating on the tip when adopting a strong graphene-tip adhesion and a normal graphene-substrate adhesion. From the load-displacement curve (blue) in Fig. 3c, we observe the extracted fitting exponent N = 1.39 and $E^* = 193\text{GPa}$, an apparent improvement compared to the model ($N = 1.30$ and $E^* = 188\text{GPa}$) when graphene is covering the substrate (black). The graphene coating on the tip provides a separation for the tip-substrate adhesion but does not affect the loading force so that in this case, graphene can be viewed as a part of the nanoindenter. The small deviation from 1.50 can also be rationalized by the influence of the normal graphene-substrate adhesion because the adhering graphene can transmit the adhesion force to the tip. However, this adhesive deviation can be substantially eliminated by applying the monolayer graphene coating on each contact side because the interactions between the graphene layers are very weak (see MD snapshots in Fig. 3c (Inset)). We also note that coating monolayer graphene directly onto the tip does not exert better effects for other three cases. However, further simulations demonstrate that coating of graphene on both sides can also positively influence the strong adhesion case regardless of the strength of the graphene-substrate adhesion (Fig. 3d). The weak tip-graphene interaction does not hamper the modification even though wrinkles emerge at the margin of graphene due to the geometrical mismatch between a sphere and a sheet (Figs. 3a-b). The full match of the lowest part of graphene and tip is sufficient for the stage of elastic contact. That is, the contact region is too shallow to touch the graphene wrinkles as shown in Figs. 3a-b (inset, red circle).

Based on these results, for strong Gr-Tip adhesion or strong Gr-Sub adhesion, the most effectual approach for monolayer graphene to diminish the adhesion effects is to apply graphene coverage to the strong-adhesion side, applying segregation to the adhesion effects. As confirmed by the simulations, in the case of strong adhesion for both graphene-tip and graphene-substrate interaction, increasing graphene layers can be an ideal method for modifying contact behavior, whereas the layer of graphene on the top layer can confine tip movement to a large extent. Nevertheless, we reveal that coating graphene on both contact sides can considerably improve the mitigation effect because the interaction between the graphene layers is quite weak. Furthermore, we must mention that the original principle of this method is similar to that in the application of graphene-coated microsphere in superlubricity reported by Liu et al. [36], in which adhesion segregation and weak interlayer forces are critical.

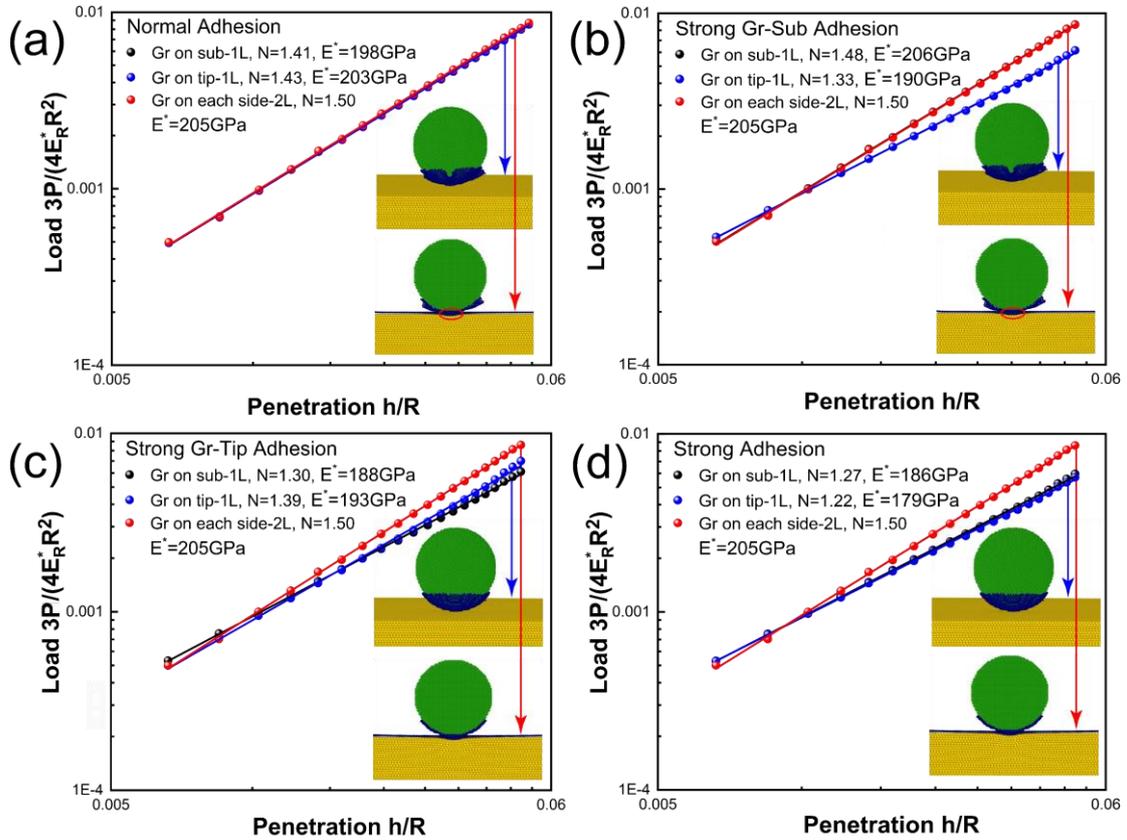

**Figure. 3.** A novel morphology of paving graphene monolayer directly onto the indenter tip for quasi adhesion-less contact for (a) normal adhesion, (b) strong Gr-Sub adhesion, (c) strong Gr-Tip adhesion, and (d) strong adhesion. Plots of force $P$ vs. scaled penetration depth $h/R$ for the graphene monolayer coating on the substrate, tip, or either side of the contacting interface are depicted in each subfigure. The insets show the snapshots of nanoindentation with graphene-monolayer-coated tips on the bare and graphene-monolayer-coated substrates.

### 3.3. Enhance substrate stiffness by applying pre-strain in graphene

Since the facial contact stiffness of the substrate is associated with the flexibility of graphene [44], we apply equivalent and incremental biaxial in-plane strain and regulate the deformability to control the adhesive contact behavior. Strain engineering is used based on our obtained results described in section 3.2 that graphene shows low surface adhesion to other materials and that atom attachment is constrained by its inherent high in-plane stiffness. Meanwhile, with regard to the friction behavior, Zhang et al. have proposed that effective modulation is achieved by tuning the flexibility of the graphene on the substrate via in-plane straining [9]. Additionally, as shown in Fig. 1d, a substrate with ultrahigh stiffness can efficiently prevent the interference from the adhesion of the substrate atoms. Therefore, a comprehensive study on the strain engineering used to tune the prediction performance is important and is carried out as described below.

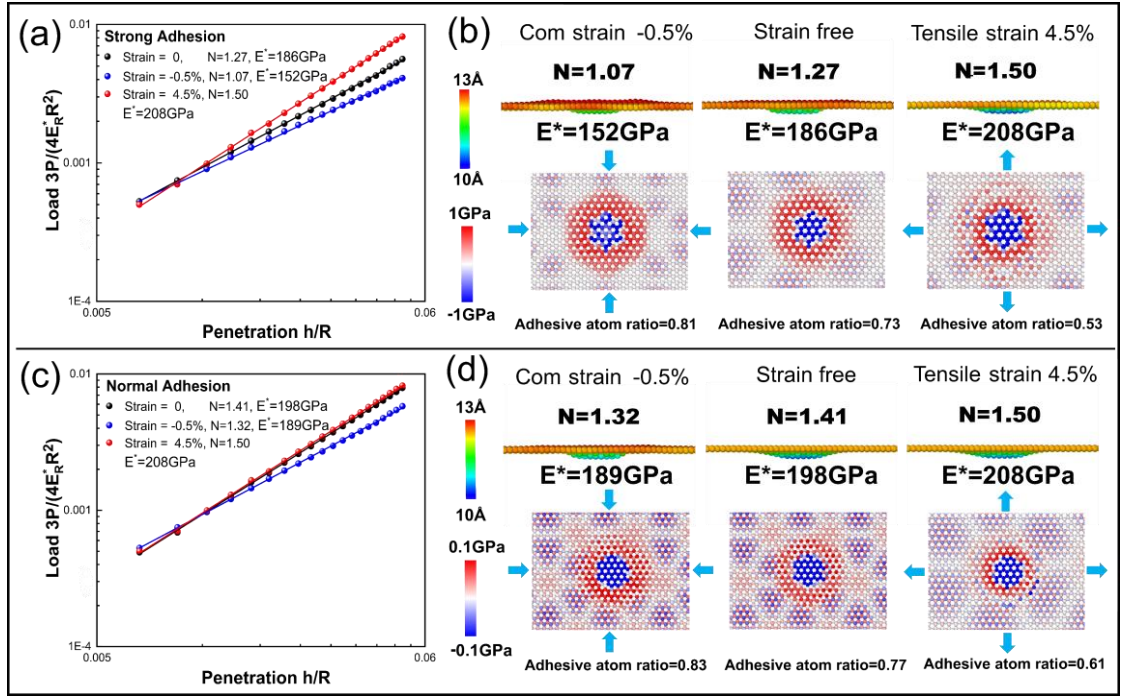

**Figure 4.** Modulation of contact behavior on the graphene-monolayer-coated Pt substrate with biaxial in-plane pre-strains for (a-b) strong adhesion and (c-d) normal adhesion. (a, c) Plots of force $P$ vs. scaled penetration depth $h/R$ for graphene monolayer with -0.5%, 0, and +4.5% biaxial pre-strain. (b, d) Side (top panel) and top (bottom panel) views for the graphene monolayer with different pre-strain when the penetration depth $h/R = 0.1$. The color for each atom represents the out-of-plane displacement ($u_z$) and atomic stress component ($\sigma_{zz}$), respectively. The ratio of adhesive atoms is calculated based on the identification of repulsive atoms ($\sigma_{zz} < 0.1\sigma_{zz}^{min}$) and adhesive atoms ($\sigma_{zz} > 0.1\sigma_{zz}^{max}$), where ($\sigma_{zz}^{min}, \sigma_{zz}^{max}$) are $(-1\text{ GPa}, 1\text{ GPa})$ and $(-0.1\text{ GPa}, 0.1\text{ GPa})$ for (b) strong adhesion and (d) normal adhesion, respectively.

In Fig. 4a, for the extreme conditions i.e., strong adhesion, compressive strain of -0.5% can noticeably enlarge the influence of adhesion on load deviations ($N = 1.07$ and $E^* = 152\text{GPa}$ for blue curve) while tensile strain up to 4.5% can completely eliminate the adhesion effects ($N = 1.50$ and $E^* = 208 \text{GPa}$ for red curve). As illustrated in the top panels of Fig. 4b, compressive strain leads to a more flexible configuration and a more intimate contact (left panel), whereas tensile strain increases the facial stiffness, producing a less-pinned graphene-tip interface (right panel). Atoms on graphene coating that provide repulsive forces prevent tip penetration, while adhesion is related to the atoms on the graphene coating that provide adhesive forces to the tip. Therefore, we calculate the normal atomic stress ($\sigma_{zz}$) distribution for the graphene monolayer at the same depth-radius ratio $h/R = 0.1$. The stress distribution of the graphene monolayer with various strains are shown in the bottom panel of Fig. 4b, indicating that the number of adhesive atoms (red, $\sigma_{zz} > 0.1 GPa$) decreases with increasing pre-strain, whereas the number of repulsive atoms (blue, $\sigma_{zz} < -0.1 GPa$) remains almost unchanged. Therefore, adhesion can be significantly mitigated by applying pre-strain in graphene prior to nanoindentation. For normal adhesion, the load-displacement curves and atomic configurations in Figs. 4c-d show exactly the same trend as those in Figs. 4a-b, indicating that the application of tensile pre-strain to the graphene coating on a substrate is an effective method for modulating the elastic behavior of nanoindentation.

To quantitatively describe the effect of in-plane strain during contact, the fitting exponent $N$ and contact modulus $E^*$ are plotted versus the biaxial strain in Fig. 5. These two fitting parameters of the load-displacement curve increase monotonically with incremental tensile strain and gradually plateau at high strain, indicating the gradual reduction and eventual elimination of adhesion interference. This outcome is understandable because strain engineering in graphene will change the out-of-plane flexibility and adjust the adhesion behavior of the graphene atoms to the tip, thereby affecting adhesion strength and contact quality. Therefore, by tuning the in-plane strain of graphene, we demonstrate that the surface adhesion of graphene-substrate system to the tip can be well regulated by altering the contact stiffness and atomic-scale adhesion quality. This conclusion provides an example of the direct regulation of the atomic-scale contact behaviors via mechanical deformation.

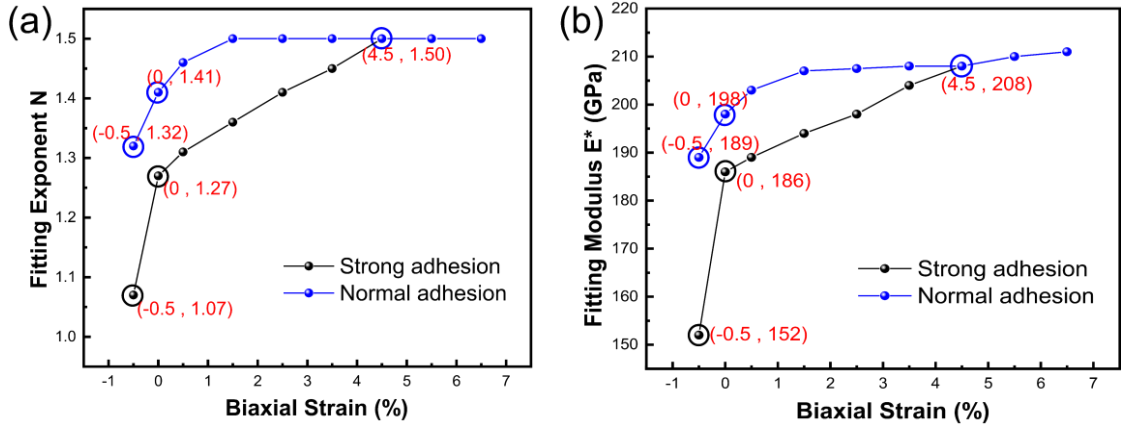

**Figure 5.** Exponent $N$ and contact modulus $E^*$ obtained by fitting $P/(4E_R^* R^2/3) = (E^*/E_R^*)(h/R)^N$ versus biaxial pre-strain of the graphene monolayer coating for strong adhesion and normal adhesion. The black and blue open circles correspond to the MS simulations shown in Figs. 4a and 4c, respectively.

### 3.4. Size effect on nanoscale adhesive contact

As represented in Fig. 6a, to explore the effect of tip size, two indenter tips with different sizes (radii of 5 and 20 nm) are produced to indent the graphene-monolayer-covered Pt substrate. For a thorough comparison, three typical tip-graphene interactions are adopted here, namely, 1) purely repulsive potential, 2) normal adhesion, and 3) strong adhesion. The values of the exponent $N$ and contact modulus $E^*$ calculating from the fittings are provided in Figs. 6b-d, showing that indentations with a larger indenter size exhibit a closer agreement with Hertz law. As illustrated in Fig. 1c (inset), the atoms providing adhesive forces to the tip (red) are only located at the margin of the contact interface, while the atoms preventing tip penetration (blue) are located on the contact surface. Hence, for the same depth-radius ratio $h/R$, we estimate that the number of adhesive atoms of graphene is proportional to the tip size, $N_a \propto R$. Based on the Hertz contact theory, the corresponding number of the repulsive atoms of graphene, i.e. the contact area, is estimated to be proportional to the square of tip size, $N_r \propto R^2$. We conclude that the proportion of adhesive atoms is inversely proportional to the tip size, $N_a/N_r \propto 1/R$, which leads to the significant influence of adhesion for smaller tip size. To further quantify the proportion of adhesive atoms for different tip sizes, we calculate the normal atomic stress ($\sigma_{zz}$) distribution for the graphene monolayer at the same depth-radius ratio $h/R = 0.2$.

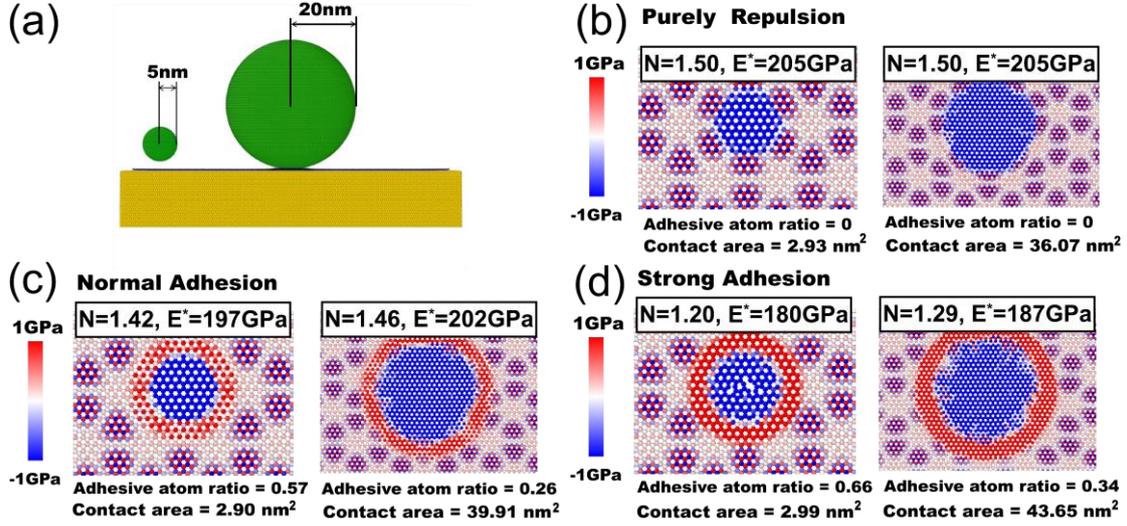

**Figure 6.** Effect of tip size on the adhesive contact behavior. (a) Schematic illustration of nanoindentation on a graphene-monolayer-covered Pt substrate with indenters of radii $R = 5$ and 20 nm. (b-d) Atomic stress distribution for the graphene monolayer on Pt substrates at the contact interface as the penetration depth $h/R = 0.2$ for 5 nm (left) and 20 nm (right) tips. The color for each atom represents the atomic stress ($\sigma_{zz}$) component. The ratio of adhesive atoms is calculated based on the identification of repulsive atoms ($\sigma_{zz} < -0.1$) GPa and adhesive atoms ($\sigma_{zz} > 0.1$)

GPa. Here three typical adhesions are applied as (b) purely repulsive interaction, (c) normal adhesion, and (d) strong adhesion. The corresponding exponents $N$ are extracted by fitting $P/(4E_R^{*}R^2/3) = (E^{*}/E_R^{*})(h/R)^N$ to the load-displacement curves from the MS simulations.

Figs. 6b-d show the stress distribution of the graphene monolayer for 5 nm and 20 nm tips for the three typical adhesion cases, respectively. It is clearly observed from Fig. 6b that a purely repulsive interaction results in entirely repulsive forces of the graphene atoms in contact, whereas the proportion of atoms with adhesive forces increases gradually with increasing adhesion as shown in Figs. 6c and 6d. The atoms with positive stress (red) provide adhesive force to indentation, whereas those with negative stress (blue) hinder tip penetration. Here, we need to point out that the periodic stress distribution out of the contact area manifests the impact of the moiré pattern formed between the graphene and the supporting crystalline substrate [45, 46]. For a more quantitative description of the contact behavior, the proportions of the atoms with adhesive forces relative to the total number of atoms in contact (adhesive atom ratio) and the contact area providing repulsive forces (repulsive contact area) are calculated for each model. Based on Figs. 6b-d, we verified our previous theoretical estimation that a larger tip possesses a lower proportion of adhesive atoms during contact. Meanwhile, for tips with a uniform size, the repulsive contact area remains constant regardless of the adhesion strength. For particular adhesion conditions, the ratios of the repulsive contact area between the tips with two different sizes are respectively, 12.31, 13.76, 14.60, close to above theoretical value (=16).

Our conclusion corroborates the results of other studies obtained by FEA method calculations [47]. It is important to note that the tip size effect is more significant for strong adhesion than for the normal adhesion, and the contact behavior of the purely repulsive case is independent of the tip size, as shown in Figs. 6b-d. In addition to the fundamental mechanism, the tip-size effect also emphasizes the significance of a better insight into the adhesive contact behavior, particularly at the nanoscale; that is, contact at this scale will strongly deviate from the traditional elastic model due to apparent adhesion.

## 4. Conclusions

In conclusion, our research demonstrates that the graphene coverage will lose its contact modification effect in some usual cases (strong Gr-Tip adhesion and compressive pre-strain of graphene), which is critical to applications. To solve this problem, we reveal that the elastic contact behavior of metallic substrate exhibits a near ideal Hertzian contact that is achieved by coating graphene layers on either side of the contact interface. This regulation effect can be explained by the high in-plane strength and low surface energy of graphene and originates from the modification of the interaction forces between the tip and the indented surface. For the bare substrate, the effect of adhesion on the nanoindentation can be significantly mitigated by reducing the amplitude and range of atomic interaction between the tip and substrate, or enhancing the substrate stiffness. This principle shows that the elastic contact behavior for specific indenter/substrate interaction can be regulated by increasing the number of coated graphene layers due to the weak interlayer interaction between the graphene layers. Furthermore, the elastic contact behavior closer to the Hertz prediction can be obtained by using a larger indenter or by applying pre-strains to graphene layers. Hence, graphene-covered nanoindentation is an effective approach to measure mechanical properties of substrate for any adhesion between tip, graphene, and substrate. Our study confirms some related experimental results reported by other researchers and provides theoretical guidance for the design of adhesion-less coatings for the AFM probes and MEMS/NEMS systems.

**Declaration of competing interest**

The authors declare that they have no known competing financial interests or personal relationships that could have appeared to influence the work reported in this paper.

**Acknowledgments**

We acknowledge support from Science Foundation of the National Key Laboratory of Science and Technology on Advanced Composites in Special Environments and the fundamental Research Funds for the Central Universities (grant no. WK2480000006).

# References:

[1]. Novoselov and S. K., Electric Field Effect in Atomically Thin Carbon Films. Science, 2004. 306(5696): p. 666-669.

[2]. Zhang, S., et al., Tribology of two-dimensional materials: From mechanisms to modulating strategies. Materials Today, 2019. 26: p. 67-86.

[3]. Berman, D., A. Erdemir and A.V. Sumant, Graphene as a protective coating and superior lubricant for electrical contacts. Applied Physics Letters, 2014. 105(23): p. 231907.

[4]. Lee, C., et al., Measurement of the Elastic Properties and Intrinsic Strength of Monolayer Graphene. Science, 2008. 321(5887): p. 385-388.

[5]. Wang, S., et al., Wettability and Surface Free Energy of Graphene Films. Langmuir, 2009. 25(18): p. 11078-11081.

[6]. Kim, K.S., et al., Chemical Vapor Deposition-Grown Graphene: The Thinnest Solid Lubricant. Acs Nano, 2011. 5(6): p. 5107-5114.

[7]. Nilsson, L., et al., Graphene Coatings: Probing the Limits of the One Atom Thick Protection Layer. ACS Nano, 2012. 6(11): p. 10258-10266.

[8]. Li, S., et al., The evolving quality of frictional contact with graphene. Nature, 2016. 539(7630): p. 541-545.

[9]. Zhang, S., et al., Tuning friction to a superlubric state via in-plane straining. Proceedings of the National Academy of Sciences, 2019. 116(49): p. 24452-24456.

[10]. Vasić, B., et al., Nanoscale wear of graphene and wear protection by graphene. Carbon, 2017. 120: p. 137-144.

[11]. Klemenz, A., et al., Contact mechanics of graphene-covered metal surfaces. Applied Physics Letters, 2018. 112(6): p. 061601.

[12]. Klemenz, A., et al., Atomic Scale Mechanisms of Friction Reduction and Wear Protection by Graphene. Nano Letters, 2014. 14(12): p. 7145-7152.

[13]. Xu, Q., et al., Suppressing Nanoscale Wear by Graphene/Graphene Interfacial Contact Architecture: A Molecular Dynamics Study. ACS Applied Materials & Interfaces, 2017. 9(46): p. 40959-40968.

[14]. Peng, W., et al., Strengthening mechanisms of graphene coatings on Cu film under nanoindentation: A molecular dynamics simulation. Applied Surface ence, 2019. 487(SEP.1): p. 22-31.

[15]. Li, Q., et al., Substrate effect on thickness-dependent friction on graphene. physica status solidi (b), 2010. 247(11-12): p. 2909-2914.

[16]. Lee, C., et al., Frictional characteristics of atomically thin sheets. Science, 2010. 328(5974): p. 76-80.

[17]. Park, S., et al., Incipient plasticity and fully plastic contact behavior of copper coated with a graphene layer. APL Materials, 2019. 7(3): p. 031106.

[18]. Hammad, M., et al., Adhesionless and near-ideal contact behavior of graphene on Cu thin film. Carbon, 2017. 122: p. 446-450.

[19]. Johnson, K.L., K. Kendall and A.D.A. Roberts, Surface Energy and the Contact of Elastic Solids. Proc.r.soc.lond.a, 1971. 324(1558): p. 301-313.

[20]. Galanov, B.A., Models of adhesive contact between rough elastic solids. International Journal of Mechanical Sciences, 2011. 53(11): p. 968-977.

[21]. Song, J. and D.J. Srolovitz, Adhesion effects in material transfer in mechanical contacts. Acta Materialia, 2006. 54(19): p. 5305-5312.

[22]. Milne, Z.B., R.A. Bernal and R.W. Carpick, Sliding History-Dependent Adhesion of Nanoscale Silicon Contacts Revealed by in Situ Transmission Electron Microscopy. Langmuir, 2019. 35(48): p. 15628-15638.

[23]. Milne, Z., et al., Covalent Bonding and Atomic-Level Plasticity Increase Adhesion in Silicon-Diamond Nanocontacts. ACS applied materials & interfaces, 2019. 11(43): p. 40734-40748.

[24]. Xu, S., et al., Molecular dynamics simulations of nano-indentation and wear of the γTi-Al alloy. Computational Materials Science, 2015. 110: p. 247-253.

[25]. He, X., Q. Bai and R. Shen, Atomistic perspective of how graphene protects metal substrate from surface damage in rough contacts. Carbon, 2018. 130: p. 672-679.

[26]. Daneshmand, M. and R.R. Mansour, RF MEMS Satellite Switch Matrices. IEEE Microwave Magazine, 2011. 12(5): p. 92-109.

[27]. Kwon, H., et al. Contact materials and reliability for high power RF-MEMS switches. 2007: IEEE.

[28]. Hyman, D. and M. Mehregany, Contact physics of gold microcontacts for MEMS switches. Components & Packaging Technologies IEEE Transactions on, 1999. 22(3): p. 357-364.

[29]. Plimpton, S., Fast Parallel Algorithms for Short-Range Molecular Dynamics. Journal of Computational Physics, 1995. 117(1): p. 1-19.

[30]. Zhou, X., R. Johnson and H. Wadley, Misfit-energy-increasing dislocations in vapor-deposited CoFe/NiFe multilayers. Physical Review B, 2004. 69(14): p. 144113-0.

[31]. Brenner, D.W., et al., A second-generation reactive empirical bond order (REBO) potential energy expression for hydrocarbons. Journal of Physics Condensed Matter, 2002. 14(4): p. 783-802.

[32]. Pastewka, L., et al., Describing bond-breaking processes by reactive potentials: Importance of an environment-dependent interaction range. Physical review B, 2008. 78(16): p. 53-56.

[33]. Wang K., et al., Improved embedded-atom model potentials of Pb at high pressure: application to investigations of plasticity and phase transition under extreme conditions, Modelling and Simulation in Materials Science and Engineering, 2018. 27(1): p 015001.

[34]. Mendelev, M.I., et al., Analysis of semi-empirical interatomic potentials appropriate for simulation of crystalline and liquid Al and Cu. Philosophical Magazine, 2008. 88(12): p. 1723-1750.

[35]. Peeters, F.M. and M. Neek-Amal, Nanoindentation of a circular sheet of bilayer graphene. Physical Review B, 2010. 81(23): p. 235421.

[36]. Liu, S.W., et al., Robust microscale superlubricity under high contact pressure enabled by graphene-coated microsphere. Nat Commun, 2017. 8: p. 14029.

[37]. Zhang, J., et al., Effects of grain boundary on wear of graphene at the nanoscale: A molecular dynamics study. Carbon, 2019. 143: p. 578-586.

[38]. Deng, Z., et al., Adhesion-dependent negative friction coefficient on chemically modified graphite at the nanoscale. Nature Materials, 2012. 11(12): p. 1032-1037.

[39]. Niu, T., G. Cao and C. Xiong, Indentation Behavior of the Stiffest Membrane Mounted on a Very Compliant Substrate: Graphene on PDMS. International Journal of Solids and Structures, 2017: p. S0020768317302512.

[40]. Zhou, L., Y. Wang and G. Cao, Estimating the elastic properties of few-layer graphene from the free-standing indentation response. J Phys Condens Matter, 2013. 25(47): p. 475301.

[41]. Suk, J.W., et al., Probing the adhesion interactions of graphene on silicon oxide by nanoindentation. Carbon, 2016. 103: p. 63-72.

[42]. Hui, F., et al., Moving graphene devices from lab to market: advanced graphene-coated nanoprobes. Nanoscale, 2016. 8(16): p. 8466-8473.

[43]. Martin-Olmos, C., et al., Graphene MEMS: AFM Probe Performance Improvement. ACS Nano, 2013. 7(5): p. 4164-4170.

[44]. Kitt, A.L., et al., How graphene slides: measurement and theory of strain-dependent frictional forces between graphene and SiO2. Nano Letters, 2013. 13(6).

[45]. Liu, J., et al., Lateral force modulation by moiré superlattice structure: Surfing on periodically undulated graphene sheets. Carbon, 2017. 125: p. 76-83.

[46]. Chan, N., et al., Contrast in nanoscale friction between rotational domains of graphene on Pt(111). Carbon, 2017. 113: p. 132-138.

[47]. Zhang, X., X. Zhang and S. Wen, Finite Element Modeling of the Nano-scale Adhesive Contact and the Geometry-based Pull-off Force. Tribology Letters, 2011. 41(1): p. 65-72.